\documentclass[wcp]{jmlr}
\usepackage{amsmath,amssymb,graphicx,url}
\jmlrvolume{152}
\jmlryear{2021}
\jmlrworkshop{Conformal and Probabilistic Prediction and Applications}
\jmlrproceedings{PMLR}{Proceedings of Machine Learning Research}

\title{Conformal Uncertainty Sets for Robust Optimization}
\author{
\Name{Chancellor Johnstone} \Email{chancellor.johnstone@afit.edu}\\
\Name{Bruce Cox} \Email{bruce.cox@afit.edu}\\
\addr{Air Force Institute of Technology, Wright-Patterson Air Force Base, OH, USA}}
\editor{Lars Carlsson, Zhiyuan Luo, Giovanni Cherubin and Khuong An Nguyen}

\usepackage{booktabs}
\usepackage{breakcites}
\usepackage{microtype}
\usepackage{xcolor}
\usepackage[load-configurations=version-1]{siunitx} 
\usepackage{float}

\theorembodyfont{\upshape}
\theoremheaderfont{\scshape}
\theorempostheader{:}
\theoremsep{\newline}

\usepackage{centernot}
\usepackage{tikz-network}
\usepackage{multicol}
\usepackage{setspace}
\usepackage{mathtools}

\mathtoolsset{showonlyrefs}

\begin{document}

\maketitle

\begin{abstract}
Decision-making under uncertainty is hugely important for any decisions sensitive to perturbations in observed data. One method of incorporating uncertainty into making optimal decisions is through robust optimization, which minimizes the worst-case scenario over some \textit{uncertainty set}. We connect conformal prediction regions to robust optimization, providing finite sample valid and conservative ellipsoidal uncertainty sets, aptly named conformal uncertainty sets. In pursuit of this connection we explicitly define Mahalanobis distance as a potential conformity score in full conformal prediction. We also compare the coverage  and optimization performance of conformal uncertainty sets, specifically generated with Mahalanobis distance, to traditional ellipsoidal uncertainty sets on a collection of simulated robust optimization examples.
\end{abstract}


\section{Introduction}
\label{sec:intro}

In many settings, the act of quantifying uncertainty related to some predicted outcome is just as important as the prediction itself. Prediction intervals allow for the attachment of bounds on random variables of interest, that tell us, with a specified probability, where observations of said random variables could fall. Ideally, we wish to construct a prediction interval $C_{1-\alpha}$ such that

\begin{equation}
    \label{eqn:pi}
    P(y \in C_{1-\alpha}) = 1 - \alpha,
\end{equation}

\noindent
where $y$ is our random variable of interest, and $\alpha$ is our desired error rate. In more complex settings, it might be of interest to capture multiple responses with a specified probability. Thus, we might construct a \textit{prediction region} such that \eqref{eqn:pi} holds for $C_{1-\alpha}\subset \mathbb{R}^d$ and some $d$-dimensioned version of $y$, say $y = (y^1, \hdots, y^d)'$. Our discussion focuses on generating finite sample valid and distribution free prediction regions through conformal prediction \citep{gammerman1998learning, vovk2005algorithmic}.

The importance of generating intervals (or regions) to accompany point predictions cannot be overstated. However, the role of these uncertainty estimates in a decision-making framework is not always explicit \citep{reckhow1994importance}. While some situations dictate an intuition-based response to uncertainty, e.g., a person avoiding an object while driving, other situations might call for a more tangible use of uncertainty estimates, e.g., assessing potential customer demand for a product \citep{bertsimas2006robust}. Multiple decision-making methods exist which explicitly take into account uncertainty, e.g., utility theory \citep{fishburn1968utility} and risk-based decision-making \citep{lounis2016risk}, but in our discussion we focus on one specific method for incorporating uncertainty into the decision-making process: robust optimization \citep{ben2009robust}. 

Robust optimization allows for the explicit utilization of uncertainty quantification in the construction of optimal and risk-averse decisions, specifically in a constrained environment. Formally, a robust optimization formulation generates an optimal solution to the problem,

\begin{equation}
    \underset{z \in \mathcal{Z}}{\min} \; \underset{u \in \mathcal{U}}{\sup}\; c(z,u),
    \label{eqn:ro-form1}
\end{equation}

\noindent
delivering some decision $z^*$ that minimizes the worst-case of some objective function $c(\cdot,\cdot)$, subject to a feasible region $\mathcal{Z}$ and an uncertainty set $\mathcal{U}$ for random parameters $u$. The uncertainty set(s) within a robust optimization formulation are meant to act as ``knob" to adjust the risk-averse nature of a decision, with larger uncertainty sets associated with a higher degree of risk-aversion. 

Given that the uncertainty set is meant to control the risk associated with a random set of parameters, their construction could benefit from the theoretical results inherent to conformal prediction. Thus, we propose a new methodological connection between conformal prediction and robust optimization. Specific contributions of the paper are listed below:

\begin{itemize}
    \item We explicitly define Mahalanobis distance as a multivariate conformity score in the full conformal prediction case. We also generalize Mahalanobis distance to be constructed in conjunction with any univariate conformity score.
    \item We introduce conformal uncertainty sets, which provide finite sample valid and distribution free uncertainty sets within the robust optimization framework.
    \item We also construct a small robust optimization example as a proof-of-concept for the use of conformal uncertainty sets, specifically with Mahalanobis distance as our conformity score of choice.
\end{itemize}

\noindent
Overall, we see that conformal uncertainty sets provide a new avenue for the construction of uncertainty sets. Additionally, we see from simulated cases that conformal uncertainty sets are competitive with traditional uncertainty sets.

In Section \ref{sec:background} we discuss relevant background to our work. Section \ref{sec:conf-uncertainty-sets} introduces the conformal approach for generating uncertainty and discusses issues that arise with full conformal prediction. Section \ref{sec:sim} relays empirical results to assess the validity, efficiency, and overall performance of conformal uncertainty sets. Section \ref{sec:conlusion} concludes the paper.

\section{Background}
\label{sec:background}

In this section we provide background on relevant topics for this paper. These topics include: conformal prediction, joint prediction regions and robust optimization.

\subsection{Conformal Prediction}
\label{sec:conf}

\noindent
Conformal prediction was first introduced in \cite{gammerman1998learning} as a method for quantifying uncertainty in both classification and regression tasks. \cite{vovk2005algorithmic} provides a formalized introduction to conformal prediction as well as application (and associated theoretical results) in multiple data settings, e.g., online and batch procedures.

We define $D_n = \{(x_i, y_i)\}_{i = 1}^n$ as a collection of $n$ observations, where the $i$-th data tuple $(x_i, y_i)$ is made up of a covariate vector $x_i$ and a response $y_i$. Our goal is to utilize $D_n$ in some fashion to construct a valid prediction interval for a new observation $(x,y)$, where $x$ is some known covariate vector and $y$ is some, yet-to-be-observed response. Assuming each data pair $(x_i, y_i)$ and $(x,y)$ are drawn exchangeably from some distribution $\mathcal{P}$, conformal prediction generates conservative, finite sample valid prediction intervals in a distribution-free manner.  Specifically, prediction intervals are generated through the repeated inversion of the test,

\begin{equation}
\begin{aligned}
&H_0: y = y_c  \\
&H_a: y \ne y_c,
\end{aligned}
\label{eqn:conf-permute}
\end{equation}

\noindent
where $y_c$ is a potential candidate response value for $y$, i.e., the null hypothesis \citep{lei2018distribution}. 

In a prediction setting, the test inversion is achieved by refitting the prediction model of interest with an augmented data set including a new data tuple $(x, y_c)$. Following the refitting, each observation in the augmented data set receives a (non)conformity score, which determines the level of (non)conformity between itself and other observations. In general, a conformity score can be any measurable function.  For example, one popular conformity score is the absolute residual

\begin{equation}
\label{eqn:abs_res}
    |y_i - \hat{y}_i(y_c)|, 
\end{equation}

\noindent
where $\hat{y}_i(y_c)$ is the predicted value for $y_i$ generated using the augmented data set. While the prediction $\hat{y}_i(y_c)$ is dependent on both $(x, y_c)$ and $D_n$, we omit dependence on $x$ and $D_n$ in our notation.

For each candidate value $y_c$ we construct $\pi(y_c)$,

\begin{equation}
\pi(y_c) = \frac{1}{n+1} + \frac{1}{n+1} \sum_{i = 1}^n \mathbb{I}\{r_i(y_c) \le r_{n+1}(y_c)\},
\label{eqn:conf-p-values}
\end{equation}

\noindent
where $r_i(y_c)$ is the conformity score associated with the data pair $(x_i, y_i)$ and $r_{n+1}(y_c)$ is the conformity score associated with $(x,y_c)$. Informally, $\pi(y_c)$ is the proportion of observations in the augmented data set whose conformity score is less than or equal to the conformity score associated with candidate value $y_c$. 

The conformal prediction region $C^{conf}_{1-\alpha}(x)$ for the incoming response $y$ is,

\begin{equation}
C^{conf}_{1-\alpha}(x) = \{y_c \in \mathbb{R} \; : \; (n+1) \pi(y_c) \le \lceil(1 - \alpha)(n+1) \rceil \}. 
\label{eqn:conf-pi}
\end{equation}

\noindent
In order to make the construction of $C^{conf}_{1-\alpha}(x)$ a tractable problem, each candidate value $y_c$ is chosen from a finite grid of points in $\mathbb{R}$, defined as the set $\mathcal{D}$. With previous results shown in \cite{vovk2005algorithmic} and \cite{lei2018distribution}, $C^{conf}_{1-\alpha}(x)$ provides a prediction interval such that

\begin{equation}
    1-\alpha \le P(y \in C^{conf}_{1-\alpha}(x)) \le 1- \alpha + \frac{1}{n+1}
\end{equation}

\noindent
for any $n$. The left-hand side of the inequality holds under exchangeability, while the right-hand side holds under the additional assumption of unique and continuous conformity scores. Thus, conformal prediction intervals are conservative, but not too conservative, while maintaining finite sample validity.



With ``full" conformal prediction, each new candidate value for the response associated with $x$ requires an additional prediction model to be fit using the augmented data set, which is computationally expensive. An alternative method, dubbed ``split" conformal prediction, utilizes an adjusted approach to full conformal prediction, but reduces the computational load required, while still achieving the same finite sample results. For clarity, we differentiate the two methods by specifically referencing them as ``full" and ``split", respectively, for the remainder for the paper.

Under the same assumptions of exchangeability, split conformal prediction only necessitates fitting the model once. The training data set $D_n$ is partitioned into two sets, $\mathcal{I}_1$ and $\mathcal{I}_2$. A prediction model is fit using $\mathcal{I}_1$ and then conformity scores are generated for each observation in $\mathcal{I}_2$. Then, the prediction interval constructed with split conformal prediction is

\begin{equation}
C^{split}_{1-\alpha}(x) = [\hat{y} - s, \hat{y} + s],
\label{eqn:split pi}
\end{equation}

\noindent
where $\hat{y}$ is the prediction for $y$ generated using the observations in $\mathcal{I}_1$, and $s$ is the $\lceil(|\mathcal{I}_2| + 1)(1-\alpha)\rceil$-th largest conformity score value for observations in $\mathcal{I}_2$.

\begin{sloppypar}
A non-exhaustive subset of CP extensions include Mondrian conformal prediction \citep{bostrom2020mondrian}, jackknife+ \citep{barber2021predictive}, aggregated conformal inference \citep{carlsson2014aggregated}, distributional conformal prediction \citep{chernozhukov2019distributional} and change detection through conformal martingales \citep{volkhonskiy2017inductive}. Advances towards conditionally valid coverage are discussed in \cite{vovk2012conditional}, \cite{guan2019conformal} and \cite{barber2019limits}, among others. For an extensive (and more recent) review of conformal prediction advances, we point the interested reader to \cite{zeni2020conformal}. 
\end{sloppypar}

\subsection{Joint Prediction Regions}
\label{sec:multi-pi}

\noindent
Instead of quantifying uncertainty for individual points, e.g., with prediction intervals, one might desire to quantify uncertainty for a collection of points. In a marginal sense, we can use simultaneous prediction intervals to, say, contain a collection of observations with some probability $1-\alpha$. Thus, we might construct an interval $C^s_{d,1-\alpha}$ such that,

\begin{equation}
    \label{eqn:spi}
    P(\textrm{all $d$ observed responses fall in $C^s_{d,1-\alpha}$}) = 1 - \alpha.
\end{equation}

\noindent
Unfortunately, generating prediction intervals for observations where the distribution is unknown is not a straightforward task. One method traditionally used to deliver conservative simultaneous prediction intervals is through the Bonferroni inequality,

\begin{equation}
\label{eqn:bonferroni}
P(\cap_{k = 1}^d \{\textrm{response $k \in C^s_{d,1-\alpha}$}\}) \ge 1 - \sum_{k = 1}^d P(\{\textrm{response $k \not\in C^s_{d,1-\alpha}$}\}),
\end{equation}

\noindent
which holds regardless of the dependency between responses $k = 1,\hdots, d$ \citep{bonferroni1936teoria}. If we construct $C^s_{d,1-\alpha}$ such that

\begin{equation*}
    P(\textrm{response $k \not\in C^s_{d,1-\alpha}$}) = \alpha^*,  
\end{equation*}

\noindent
where $\alpha^* = \frac{\alpha}{d}$, then $P(\cap_{k = 1}^d \{\textrm{response $k \in C^s_{d,1-\alpha}$}\}) \ge 1 - \alpha$ and we have a conservative, simultaneous prediction interval.

Obviously, the construction of a single prediction interval to contain a collection of points only makes sense if the observations are identically distributed. In the case of multi-target regression, where we are interested in multiple responses, this is not necessarily the case. Thus, it would be beneficial to construct a prediction region to quantify uncertainty associated with a collection of responses jointly, rather than marginally.

While many methods exist for generating prediction regions, e.g., bootstrap prediction regions \citep{beran1992designing} and Bayesian prediction regions \citep{datta2000bayesian}, our work builds from results in \cite{scheffe1959analysis}, which introduces a prediction region $C_{d,1-\alpha}$ for a $d \times 1$ vector of responses $y$,

\begin{equation*}
\label{eqn:scheffe}
    C_{d,1-\alpha} = \{y: (y - \mu)'\Sigma^{-}(y - \mu) \le \chi^2_{d, 1-\alpha}\},
\end{equation*}

\noindent
where $\mu$ is the mean of the random vector $y$, $\Sigma^{-}$ is the inverse-covariance matrix of $y$ and $\chi^2_{d,\tau}$ is the $\tau$-th quantile associated with the $\chi^2$ distribution with $d$ degrees of freedom. In practice, the true mean and covariance are usually unknown. Thus, we replace $\mu$ and $\Sigma^{-}$ with their estimates, $\hat{\mu}$ and $\hat{\Sigma}^{-}$, respectively, resulting in a prediction region

\begin{equation}
\label{eqn:scheffe-est}
    C_{d,1-\alpha} = \{y: (y - \hat{\mu})'\hat{\Sigma}^{-}(y - \hat{\mu}) \le \chi^2_{d, 1-\alpha}\},
\end{equation}

\noindent
which is asymptotically valid. Our focus on generating conformal prediction regions in a similar fashion is purely pragmatic as it supports the connection of conformal prediction regions to robust optimization, which we provide background on in Section \ref{sec:robust}.


The first results extending conformal prediction to the multivariate response case (to our knowledge) come from \cite{lei2015conformal}, which applies conformal prediction to functional data, providing bounds associated with prediction ``bands". \cite{diquigiovanni2021conformal} extends and generalizes additional results for conformal prediction on functional data. 

The functional case is inherently multidimensional and can be applied to a prediction region setting, but more focused work on prediction regions also exists. Joint conformal prediction regions were explicitly introduced in \cite{kuleshov2018conformal} and \cite{neeven2018conformal}. More recent advances include \cite{messoudi2020conformal}, which delivers Bonferroni-type prediction intervals through neural networks. \cite{messoudi2021copula} extends this work through the use of copula-based conformal prediction. \cite{cella2020valid} use Tukey's half-space depth \citep{tukey1975mathematics} as a conformity score for the multivariate response case. \cite{kuchibhotla2020exchangeability} explicitly describes conformal regions for arbitrary spaces, delivering theoretical results to support the validity of conformal prediction regions in $\mathbb{R}^d$. \cite{chernozhukov2018exact} provides valid conformal prediction regions for potentially dependent data, generalizing the traditional conformal inference assumption of exchangeability.

\subsection{Robust Optimization}
\label{sec:robust}

While real-world decision-making requires simplifying assumptions, the assumption of data certainty can be a damaging one. When constructing a formulation for an optimization problem, parameters in the formulation might come from subject matter expertise, historical data or even physical limitations. However, small perturbations in these values might not only result in the current solution being suboptimal, but also potentially infeasible \citep{ben2002robust}. 

Stochastic optimization \citep{shapiro2014lectures} aims to take uncertainties into account within a decision-making process. For some set of decision variables $z$, a stochastic optimization problem might take the form,

\begin{equation}
    \underset{z \in \mathcal{Z}}{\min} \; \textrm{E}[c(z,u)],
    \label{eqn:stoch-form}
\end{equation}

\noindent
where $\mathcal{Z}$ is the feasible region, $c(\cdot, \cdot)$ is the objective function, and $u = (u_1,\hdots,u_d)'$ is a $d \times 1$ vector of random parameters we do not observe until our decision timeline is over. While not denoted explicitly, the feasible region $\mathcal{Z}$ might also depend on $u$. For our discussion, we assume $\mathcal{Z}$ is independent of the random parameters. 



Under the stochastic optimization framework, there are multiple approaches to address uncertainty. A subset includes chance-constrained programming \citep{charnes1959chance,miller1965chance}, sample-average approximation \citep{kleywegt2002sample} and robust optimization \citep{ben2009robust, bertsimas2011theory}. We limit further discussion to the latter.




In order to incorporate uncertainty into the optimization framework, robust optimization (RO) utilizes \textit{uncertainty sets} on the random parameters, solving the problem, 

\begin{equation}
    \underset{z \in \mathcal{Z}}{\min} \; \underset{u \in \mathcal{U}}{\sup}\; c(z,u).
    \label{eqn:ro-form}
\end{equation}

\noindent
Thus, RO minimizes the potential worst-case outcome with respect to the uncertainty set $\mathcal{U}$. Uncertainty sets can be constructed in either a naive (without data), or a data-driven manner. Without observations of the random parameters we can assume interval constraints associated with each element of $u$, constructing an uncertainty set $\mathcal{U}_{box}$,

\begin{equation}
    \mathcal{U}_{box} = \{u \in \mathbb{R}^d: u^j \in [\underline{a_j}, \overline{b_j}] \; \forall \; j = 1, \hdots, d \},
    \label{eqn:Ubox}
\end{equation}

\noindent
with some lower and upper endpoints, $\underline{a_j}$ and $\overline{b_j}$, respectively. Barring pathological cases, the use of $\mathcal{U}_{box}$ in convex optimization results in each of the uncertain elements obtaining some boundary value. Given a historical data set of observed uncertain parameters we can also provide data-driven, norm-based uncertainty sets, 

\begin{equation}
    \mathcal{U}_{q} = \{u \in \mathbb{R}^d: ||u - \bar{u}||_q \le \Omega\},
    \label{eqn:Unorm}
\end{equation}

\noindent
where $||\cdot||_q$ is the $q$-norm, $\bar{u}$ is the vector of sample means associated with observations of the $d$ uncertain parameters, and $\Omega$ is some \textit{budget of uncertainty}. \cite{bertsimas2006robust} highlight that data-driven uncertainty sets like \eqref{eqn:Unorm} outperform ``naive" uncertainty sets like \eqref{eqn:Ubox}. Selecting $\mathcal{U}_q$ with $q = 2$ results in a strictly spherical uncertainty set. Adjusting the errors $u - \bar{u}$ by some matrix $A$ allows for the construction of ellipsoidal uncertainty sets,

\begin{equation}
    \mathcal{U}_{ell} = \{u \in \mathbb{R}^d: ||A(u - \bar{u})||_2 \le \Omega\}.
    \label{eqn:Uell}
\end{equation}

\noindent
For our discussion, we focus our discussion on ellipsoidal uncertainty sets. 

\begin{figure}[h]
    \centering
    \makebox[\textwidth][c]{
    \includegraphics[scale = .7, trim = 125 25 180 75, clip]{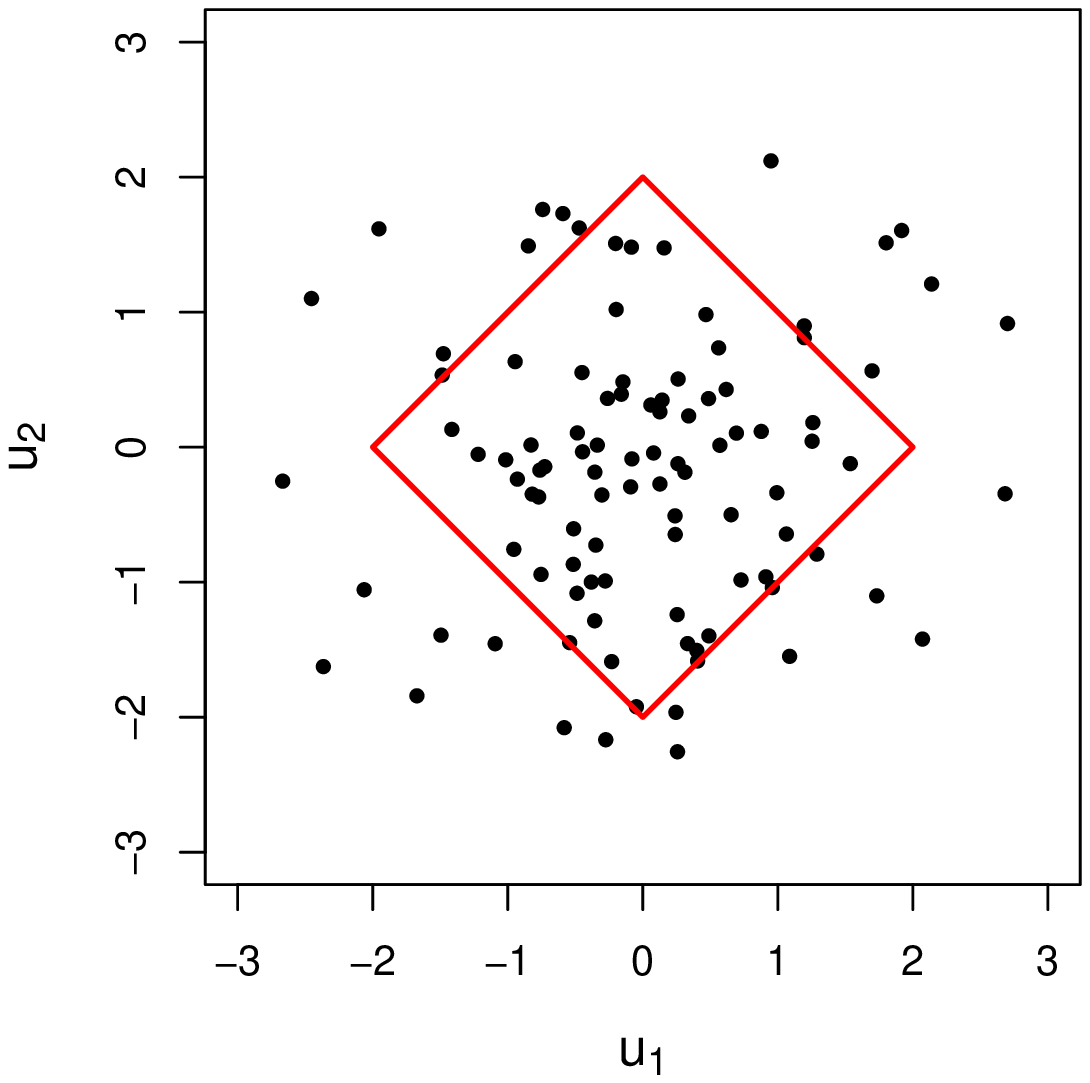}
    \includegraphics[scale = .7, trim = 125 25 180 75, clip]{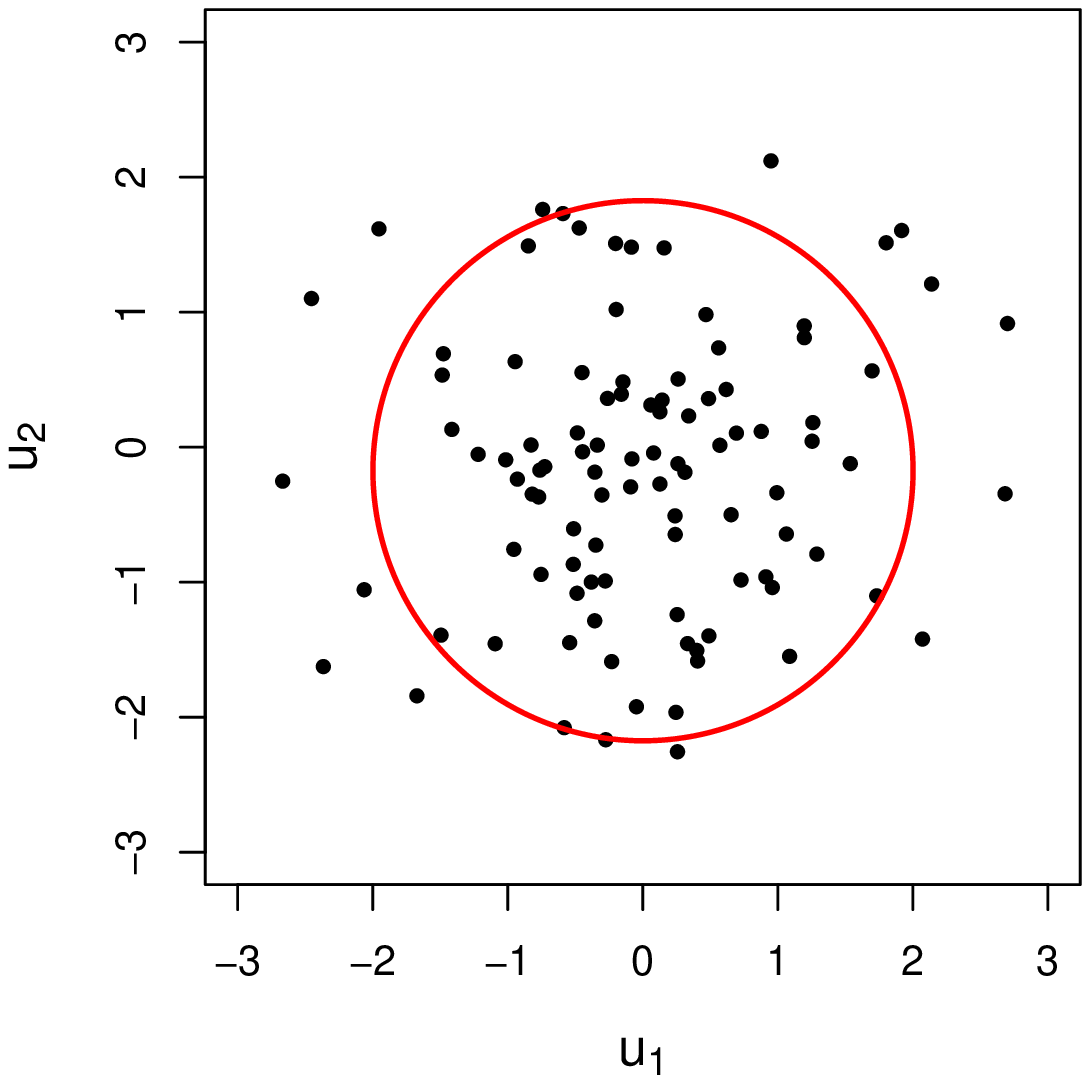}
    }
    \caption{Uncertainty set constructed with $q = 1$ (left) and $q = 2$ (right) for random parameters $u = (u_1, u_2)'$ with $\Omega = 2$. Randomly generated observations of $u$ shown as ``$\bullet$".}
    \label{fig:Usets}
\end{figure}

A larger value of $\Omega$ corresponds to a more risk-averse decision. Thus, a trade-off exists between the optimal solution and the ``robustness" of said solution. Figure \ref{fig:Usets} shows examples of norm-based uncertainty sets constructed with different values for $q$ for a two-dimensional data set drawn from two independent standard normal random variables.

To solidify the intuition behind robust optimization we construct a toy example. Suppose we have the vector of decision variables $z = (z_1, z_2)'$ and a vector of random parameters $u = (u_1, u_2)'$. In this case our random parameter vector is some cost and our goal is to minimize said cost. For this example, we assume $u$ to be multivariate normal with mean equal to the zero vector and covariance matrix with marginal variance of one and correlation of $0.5$. The robust optimization formulation of interest is,



\begin{equation}
\label{eqn:ro-toy}
\begin{array}{ll@{}l}
\underset{z}{\min} \; \underset{u \in \mathcal{U}}{\sup}\;  & \displaystyle\sum\limits_{k=1}^{2} u_{k}z_{k}&\\
\text{subject to}& z_1 + z_2 = 1\\
                 & 0 \le z_{k} \le 1 & \; \forall \; k =1,2
\end{array}
\end{equation}

\noindent
where $\mathcal{U} = \{u \in \mathbb{R}^2: ||L^-(u - \mu)||_2 \le 2\}$ and  $L^-$ is the lower triangular matrix of the Cholesky decomposition for the inverse-covariance matrix associated with $u$.

\begin{figure}[h]
    \centering
    \makebox[\textwidth][c]{
    \includegraphics[scale = .65, trim = 0 15 0 70, clip]{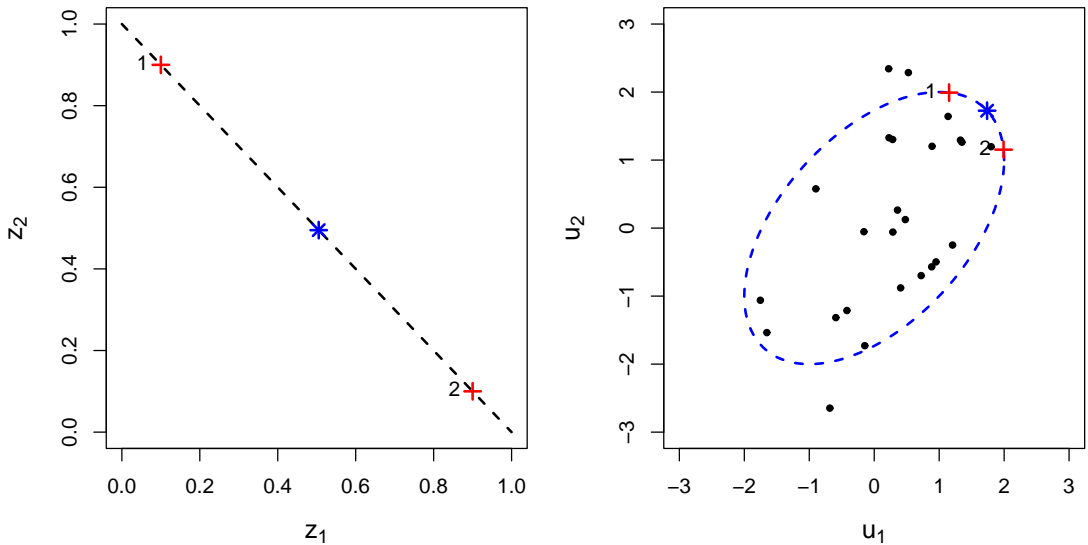}
    }
    \caption{Illustration of \eqref{eqn:ro-toy}, showing the feasible region (left) and uncertainty set (right). Sub-optimal decision variables and associated worst-case realizations of random parameters are labelled with ``\textcolor{red}{+}" and numbered accordingly. Optimal solution and minimal worst-case realization labeled with ``\textcolor{blue}{\textbf{$\ast$}}".}
    \label{fig:robust-toy-illus}
\end{figure}

The intuition behind \eqref{eqn:ro-toy} is illustrated in Figure \ref{fig:robust-toy-illus}. In the left plot we show the feasible region for $z$ as the dotted line from $(0,1)$ to $(1,0)$. Along the feasible region we identify two sub-optimal feasible solutions at $(0.1,0.9)$ and $(0.9, 0.1)$ with red plusses. In the right plot we identify the border of the uncertainty set $\mathcal{U}$ with the blue dotted line. We also denote the worst-case value of $u$ for each sub-optimal feasible solution identified in the left plot, $u = (1.15, 1.99)'$ and $u = (1.99, 1.15)'$ respectively, with red plusses. Twenty-five sample observations of $u$ are also shown. The optimal solution at $z^* = (0.5, 0.5)'$, along with the corresponding worst-case realization, $u = (1.73, 1.73)'$, are identified with blue asterisks.

Even though most robust optimization problems are inherently more difficult to solve than, say, linear optimization problems, tractable formulations do exist. Intimate details of of these tractable implementations are not crucial to our discussion. Additional insight on robust optimization as well as practical applications can be seen  in \cite{bertsimas2011theory} and  \cite{gorissen2015practical}, among others.








\section{Conformal Uncertainty Sets for Robust Optimization}
\label{sec:conf-uncertainty-sets}

Other works, e.g., \cite{bertsimas2018data}, \cite{delage2010distributionally}, \cite{chen2010cvar} and \cite{lu2011robust}, have reduced the distributional assumptions required to generate uncertainty sets with better probabalistic guarantees, but these results still rely on distributional assumptions of the data and/or on some large sample requirement.

In this section we explore Mahalanobis distance as a conformity score for multivariate responses and connect to uncertainty sets for robust optimization. For completeness, we define and generalize Mahalanobis distance for use in full conformal prediction. In Section \ref{sec:conf-uncertainty-sets2}, we shift our focus to split conformal prediction.
 
\subsection{Conformal Prediction with Mahalanobis Distance}
\label{sec:conf-us}

Instead of making assumptions on the normality of our variables of interest and/or if we have enough observations for large sample approximations, we can use conformal prediction to generate finite sample, distribution free prediction regions for any set of multivariate random variables.

\subsubsection{Full Conformal Approach}

Suppose again we define $D_n = \{(x_i, y_i)\}_{i = 1}^n$ as a collection of $n$ observations. In this case, the $i$-th data tuple $(x_i, y_i)$ is made up of a covariate vector $x_i$ and a response vector $y_i = (y^1_i,\hdots,y^d_i)'$, which slightly differs from the tuple construction in Section \ref{sec:conf}. We also assume each data tuple $(x_i, y_i)$ and our new observation $(x,y)$, where $y = (y^1,\hdots,y^d)'$, are drawn exchangeably from some distribution $\mathcal{P}^d$.

The prediction region of interest is one which contains the new response $y$ with probability $1 - \alpha$. We define a candidate value vector $y_c = (y^1_c, \hdots, y^d_c)'$ and $\hat{y}^k_i(y^k_c)$ as the estimate of $y^k_i$, constructed using the augmented data set. Instead of a univariate conformity score, e.g., the absolute residual for a candidate response value, we can construct a multivariate conformity score $\sigma_i(y_c)$,

\begin{equation}
\label{eqn:multi-conf}
    \sigma_i(y_c) = \sqrt{r_i(y_c)'\hat{\Sigma}^{-}r_i(y_c)},
\end{equation}

\noindent
where $r_i(y_c) = (r^1_i(y^1_c), \hdots, r^d_i(y^d_c))'$, $r^k_i(y^k_c)$ is a univariate conformity score associated with the $k$-th response of $(x_i, y_i)$, and $\hat{\Sigma}^{-}$ is the sample inverse-covariance matrix associated with the univariate conformity scores. We ignore the dependence of $\hat{\Sigma}^{-}$ on $y_c$  in notation. We explicitly define $\hat{y}_i(y_c) = (\hat{y}^1_i(y_c), \hdots, \hat{y}^d_i(y_c))'$ as the vector of $d$ individual predictions for each element in $y_i$. In our case, for a given observation $i = 1,\hdots, n$ we specify $r^k_i(y_c) = y^k_i - \hat{y}^k_i(y_c^k)$, the residual associated with the $k$-th response. However, $r^k_i(y)$ can be generally defined as any univariate conformity score. We define $\sigma(y_c)$, with no indexing, as the conformity score associated with $(x, y_c)$.

For univariate conformal prediction we utilize a set of candidate values $\mathcal{D}$ for each new observation $x$. The multivariate case instead requires a set of candidate values $\mathcal{D}^d \subset \mathbb{R}^d$. Analogous to \eqref{eqn:conf-p-values}, we define a new version of $\pi(\cdot)$ utilizing \eqref{eqn:multi-conf},

\begin{equation}
\pi(y) = \frac{1}{n+1} + \frac{1}{n+1} \sum_{i = 1}^n \mathbb{I}\{\sigma_i(y) \le \sigma(y_c)\}.
\label{eqn:mult_conf-p-values}
\end{equation}

\noindent
Then, a conservative $100(1-\alpha)\%$ conformal prediction region is formed by

\begin{equation}
C^{conf}_{d,1-\alpha}(x) = \{y \in \mathbb{R}^d \; : \; (n+1) \pi(y) \le \lceil(1 - \alpha)(n+1)\rceil \}.
\label{eqn:pred-region}
\end{equation} 

\noindent
With $r^k_i(y^k_c)$ as defined previously, \eqref{eqn:multi-conf} is the Mahalanobis distance between the vector $y_i$ and $\hat{y}_i(y_c)$. Thus, the conformal predictions regions generated are analogous to \eqref{eqn:scheffe-est}. A similar version of \eqref{eqn:multi-conf} was suggested as a conformity measure for split conformal prediction in \cite{kuchibhotla2020exchangeability}. Its explicit definition in full conformal prediction is new, but also brings an increased computational burden to its use.


While it is not obvious, prediction regions generated with the full conformal approach are not necessarily ellipsoidal (or convex), which can can problems from an optimization perspective. \cite{ben2009robust} show that in a convex optimization problem, we can replace any uncertainty set with its convex hull, but we do not explore this further in our discussion.

\begin{figure}[t]
    \centering
    \makebox[\textwidth][c]{
    \includegraphics[scale= .65, trim = 120 0 150 50, clip]{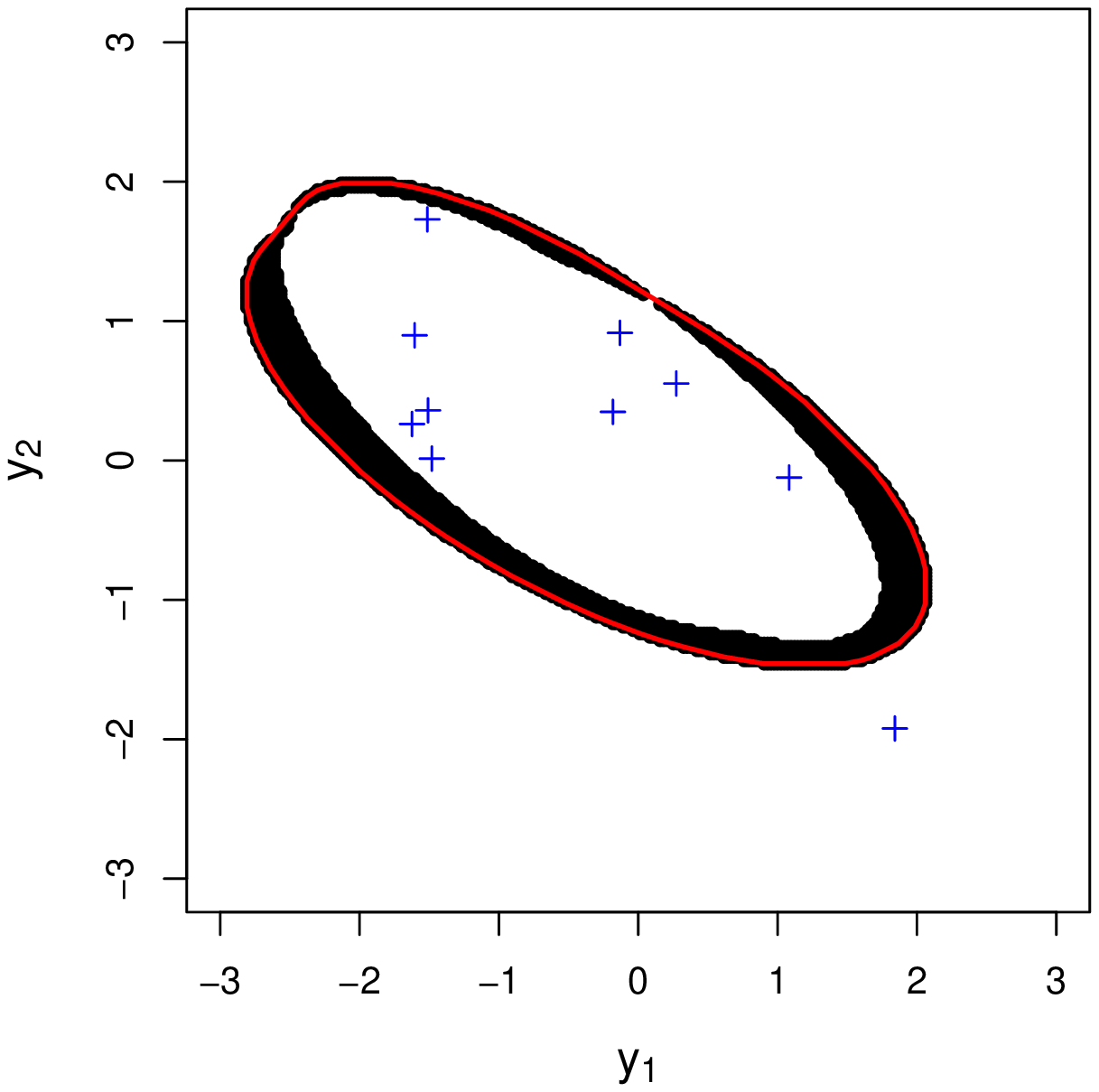}%
    \includegraphics[scale= .65, trim = 120 0 150 50, clip]{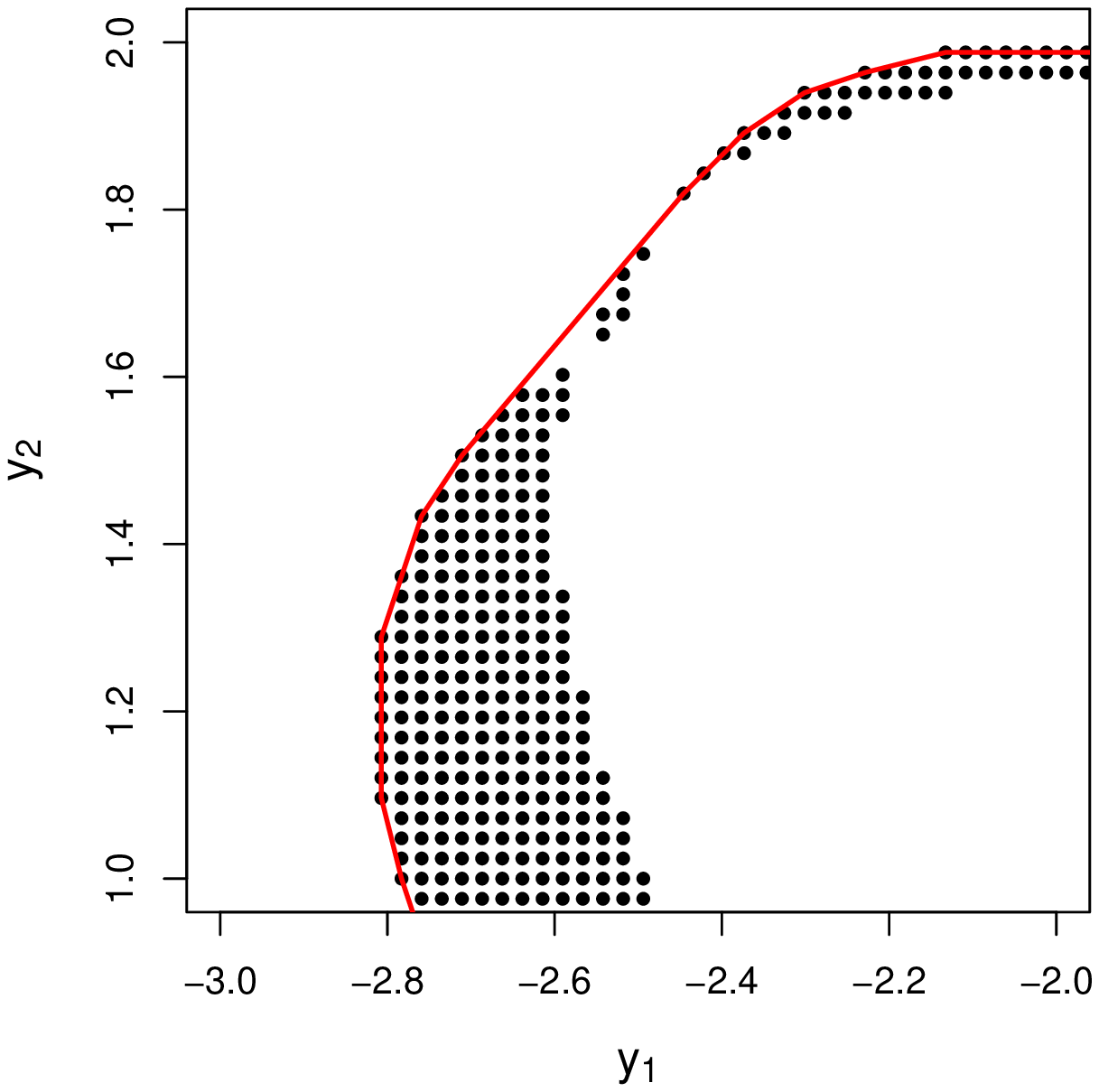}
    }
    \caption{Example of nonconvexity for full conformal prediction region with sample covariance of univariate conformity scores. Candidate value vectors such that $(n+1) \pi(y) = \lceil(1 - \alpha)(n+1)\rceil$ are shown as ``$\bullet$", which constructs the prediction region border. The convex hull of the candidate points on the border region are shown in red. Random observations of $y = (y_1, y_2)'$ are denoted with ``\textcolor{blue}{+}". We show the entire set of border points for the full conformal prediction region (left) and a zoomed-in version (right) to highlight nonconvexity.}
    \label{fig:non-convex}
\end{figure}


\subsubsection{Split Conformal Approach}

The conformity score identified in \eqref{eqn:multi-conf} can be extended to the split conformal approach. However, careful consideration must be taken when constructing $\hat{\Sigma}^{-}$. Given that split conformal requires the partitioning of the data set $D_n$ into $\mathcal{I}_1$and $\mathcal{I}_2$, \cite{kuchibhotla2020exchangeability} suggested the use of $\hat{\Sigma}^{-}_{\mathcal{I}_1}$, an estimate of the univariate conformity score inverse-covariance matrix generated from observations in $\mathcal{I}_1$.

With the split conformal procedure, we can generate a closed-form for our split conformal prediction region by using $s$, the $\lceil(1 - \alpha)(|\mathcal{I}_2|+1)\rceil$ largest value of $\{\sigma_i(y_c): i \in \mathcal{I}_2\}$. With \eqref{eqn:multi-conf} as our conformity score, we generate a prediction region,

\begin{equation}
C^{split}_{d,1-\alpha}(x) = \{y \in \mathbb{R}^d: \sigma(y) \le s\},
\label{eqn:split pred-region}
\end{equation} 

\noindent
which is the split conformal equivalent of \eqref{eqn:pred-region}. Because $\sigma(\cdot)$ is quadratic, $C^{split}_{d,1-\alpha}(x)$ is an ellipsoidal prediction region, with $s$ defining the ellipsoid border. Thus, we can construct the ellipsoidal prediction region by finding each $(x,y_c)$ tuple such that the equality portion of \eqref{eqn:split pred-region} holds. This differs from the univariate split conformal case, where the prediction intervals are defined explicitly by observations in $\mathcal{I}_2$.

Due to the potential lack of convexity for full conformal prediction regions constructed with Mahalanobis distance, we utilize the split conformal results for extension to uncertainty sets in Section \ref{sec:conf-uncertainty-sets2}. 




\subsection{Extension to Uncertainty Sets}
\label{sec:conf-uncertainty-sets2}

For RO, the $\Omega$ parameter in \eqref{eqn:Unorm} and \eqref{eqn:Uell} can be chosen by a decision maker, with higher values of $\Omega$ corresponding to higher levels of risk aversion. However, we can also select $\Omega$ in such a way as to obtain probabilistic guarantees associated with the uncertainty set constructed, e.g., through a joint prediction region constructed under normality. With reasonable, i.e., consistent, estimates of the true mean and true covariance matrix for $u$ and $\Omega$ selected as the square root of $\chi^2_{d, 1-\alpha}$, \eqref{eqn:scheffe-est} provides an asymptotically valid ellipsoidal uncertainty set, as long as normality holds.

In contrast to ellipsoidal uncertainty sets constructed under normality, we can also generate uncertainty sets with probabalistic bounds through split conformal prediction in a distribution free manner. While the terms ``prediction region" and ``uncertainty set" are normally used in the context of uncertainty quantification and robust optimization, respectively, they become equivalent in the case of conformal uncertainty sets.

By the results previously discussed, we can provide a finite sample, distribution free prediction region for $u$ through split conformal prediction. We can then use this region as a conservative uncertainty set in any robust optimization problem. Specifically, we select $\Omega$ such that $\lceil(|\mathcal{I}_2| + 1)(1-\alpha)\rceil$ of the conformity scores constructed from $\mathcal{I}_2$ are less than or equal to $\Omega$, guaranteeing a valid uncertainty set. We formalize the selection of $\Omega$ for some fixed $\alpha$ as,

\begin{equation}
\Omega_{\alpha} = \underset{\sigma_i(y)}{\sup} \{(x_i,y_i) \in \mathcal{I}_2: (|\mathcal{I}_2| + 1)\pi(y) \le \lceil(|\mathcal{I}_2| + 1)(1-\alpha)\rceil\}  
\label{eqn:omega-alpha}
\end{equation}

\noindent
where $\sigma_i(\cdot)$ is constructed using \eqref{eqn:multi-conf}. We require the supremum over $\sigma_i(\cdot)$ because we are looking for the specific value associated with the border of the conformal prediction region, giving us the effective radius of our ellipsoid, which aligns exactly with the prediction region constructed in \eqref{eqn:split pred-region}. Using $\Omega_{\alpha}$ as constructed with \eqref{eqn:omega-alpha} allows for ellipsoidal uncertainty sets of the form,

\begin{equation}
\label{eqn:conf-u}
    \mathcal{U} = \{u \in \mathbb{R}^d: ||\hat{L}_{\mathcal{I}_2}^{-}(u - \bar{u})||_2 \le \Omega_{\alpha}\},
\end{equation}

\noindent
where $\hat{L}^-_{\mathcal{I}_2}$ is the lower triangular matrix of the Cholesky decomposition for the inverse-covariance matrix associated with the conformity scores of observations in $\mathcal{I}_2$.
Uncertainty sets of the form in \eqref{eqn:conf-u} are easily implementable in currently-existing robust optimization solvers. Because \eqref{eqn:conf-u} is constructed through split conformal prediction, we achieve the finite sample results associated with conformal prediction and can generate finite sample valid, distribution free uncertainty sets for robust optimization.



\section{Performance Comparison}
\label{sec:sim}

We compare the efficiency, coverage and performance of conformal uncertainty sets to traditional ellipsoidal uncertainty sets in a simulated setting, generalizing \eqref{eqn:ro-toy} to $d$ dimensions,

\begin{equation}
\label{eqn:ro-toy-d}
\begin{array}{ll@{}l}
\underset{z}{\min} \; \underset{u \in \mathcal{U}}{\sup}\;  & \displaystyle\sum\limits_{k=1}^{d} u_{k}z_{k}&\\
\text{subject to}& \displaystyle\sum\limits_{k=1}^{d} z_{k} = 1\\
                & 0 \le z_{k} \le 1 & \; \forall \; k =1, \hdots, d.
\end{array}
\end{equation}

\noindent

\noindent
Three different uncertainty sets are considered: no uncertainty, ellipsoidal uncertainty under normality, and conformal ellipsoidal uncertainty. We construct a set of experiments by varying data generation schemes, the number of historical observations of the random parameter vector $u$, the significance level $\alpha$, and the dimension $d$. The different data generation schemes include:

\begin{itemize}
    \item both normally distributed, and $t$-distributed costs $u_k$
    \item various random parameter mean-variance  scenarios: 1) independent and identically distributed, 2) independent, but not identically distributed, and 3) correlated 
\end{itemize}

\noindent
We generate correlated, marginally $t$-distributed random parameters through the probability integral transform of multivariate normal random variable.


For every combination of the data generation schemes above we perform experiments with varying values for $\alpha = 0.05, 0.1, 0.25, 0.5$, $n = 100, 250, 1000, 2500$ and $d = 2, 10$. For each instance of an experiment we:
\begin{enumerate}
\item Generate both training and test data according to that experiments data generation scheme and set-up (i.e., choice for $\alpha$, $n$, and $d$).
\item Using the training data we construct our uncertainty sets, and solve the robust optimization model given this training data. 
\item The solution of this problem provides an optimal decision variable w.r.t. the training data. We then calculate objective function values, using this optimal decision, across the \emph{test data set}. This generates a distribution of objective function values (one for each entry in the test data set). 
\item We then calculate the $1-\alpha$ quantile of this distribution and store.
\item We repeat each experiment 100 times, recording the $\alpha$ worst-case costs for each optimal decision on a test set of 250 random parameter vectors.
\end{enumerate}      

\begin{figure}
    \centering
    \resizebox{\textwidth}{!}{
    \includegraphics{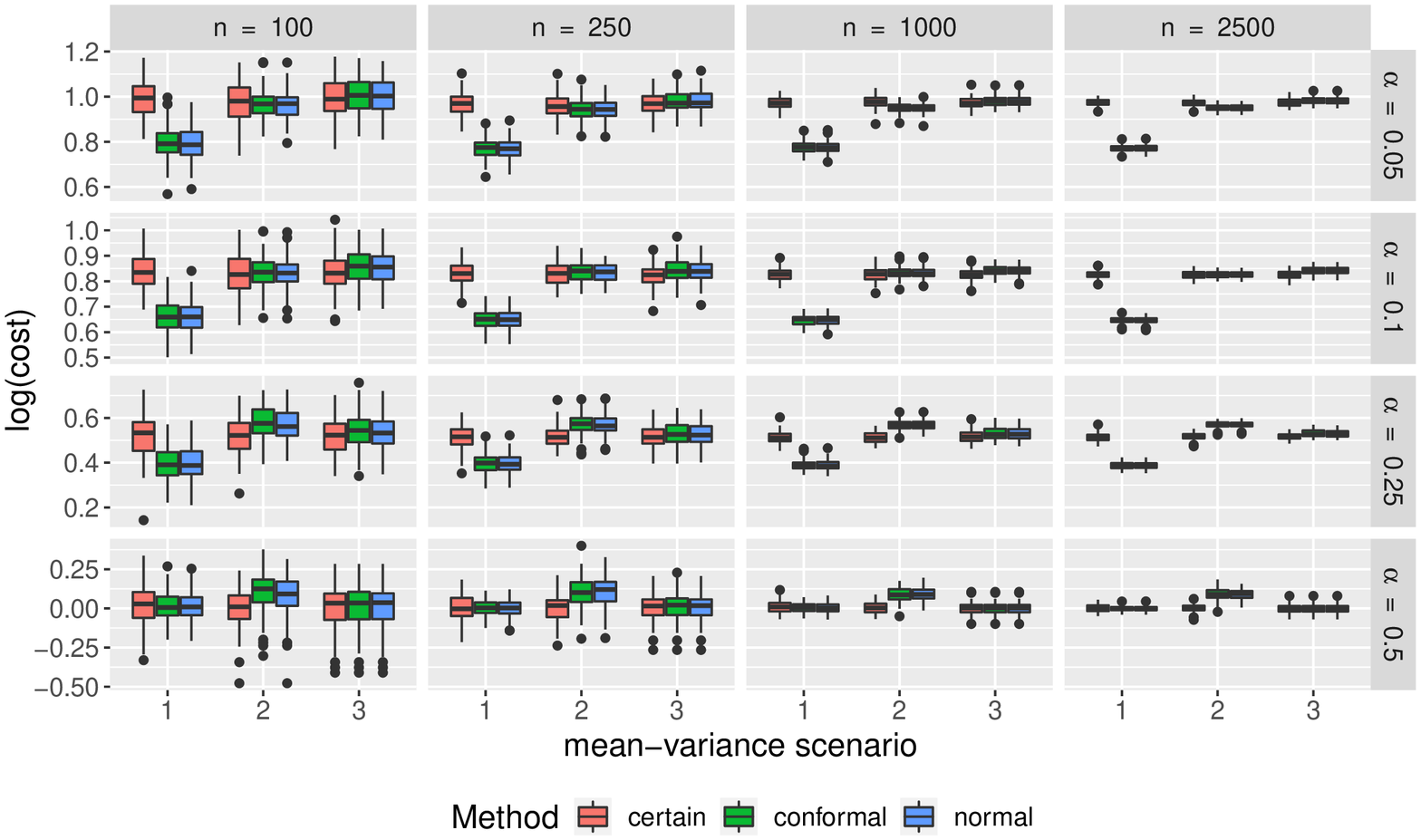}
    }
    \resizebox{\textwidth}{!}{
    \includegraphics{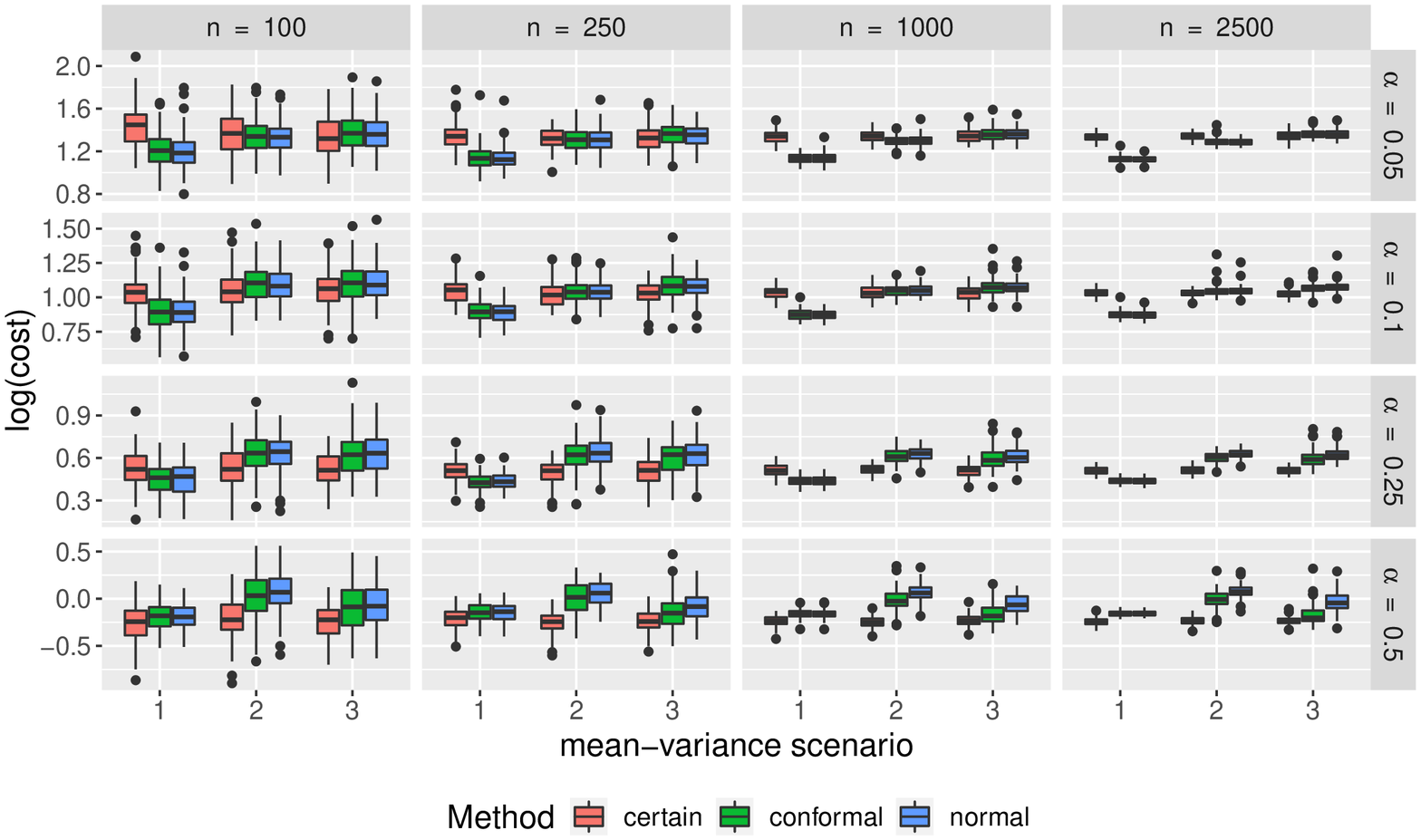}
    }
    \caption{$\alpha$ worst-case objective function value across $n$ test cases with 100 experiment repetitions for normally distributed cost (top) and $t$-distributed cost (bottom) with $d = 2$.}
    \label{fig:costd2}
\end{figure}

Figure \ref{fig:costd2} compares the overall robust optimization performance of the three uncertainty sets of interest in terms of $\alpha$ worst-case performance. We see that while the conformal uncertainty sets results in $\alpha$ worst-case scenarios comparable to uncertainty sets constructed under normality, they do not seem to be superior. We also see under mean-variance scenarios one and two, the non-robust optimization has better worst-case scenario performance than any of the robust formulations.

Comparisons of empirical coverage for each uncertainty set are shown in Supplementary Materials. We see the coverage for the conformal uncertainty sets is much closer to nominal in both the cases of normally distributed and $t$-distributed random parameter vectors. While not shown, we also find conformal uncertainty sets to be more efficient than those constructed under normality. 




\section{Conclusion}
\label{sec:conlusion}

In this paper we showed empirically the effectiveness of the Mahalanobis distance as a multivariate conformity score, specifically for the use in the generation of uncertainty sets for robust optimization. Conformal uncertainty sets not only provide finite sample validity, but are also empirically more efficient than uncertainty sets based on normality. However, more exploration is needed to assess the poor performance from a robust optimization perspective. For prediction regions, it is widely accepted that without validity, efficiency is not informative. Thus, comparing the $\alpha$ worst-case performance of the uncertainty sets constructed when the coverage of the normality-based prediction regions is not valid, let alone conservative, is a potentially unfair one. Our comparisons also fail to describe the potential trade-off between robustness and average performance. Thus, further exploration is needed to see where conformal uncertainty sets might better perform. Additionally, in the future we hope to explore performance in a practical setting, rather than just a simulated one.

In our discussion we chose Mahalanobis distance as our sole conformity score due to the inherent relationship between itself and traditional ellipsoidal uncertainty sets. However, another conformity score, referenced in Section \ref{sec:conf}, is half-space depth. With half-space depth we can decompose a set of points in $\mathbb{R}^d$ into a nested set of convex regions \citep{bremner2007halfspace}, which would be helpful in the construction of \textit{polyhedral} uncertainty sets. Future work might compare the performance of conformal uncertainty sets constructed with Mahalanobis distance and half-space depth.

We focused on a robust optimization problem where there was uncertainty only in the objective function coefficients. Applying conformal uncertainty sets to more complex robust optimization formulations, e.g., with uncertainty in the constraints, is a natural extension. Additionally, utilizing conformal prediction regions to provide probabalistic controls on constraint violation would also be useful. We also assumed independence of the feasible region $\mathcal{Z}$ and uncertain parameters $y$. Relaxation of this assumption could be beneficial as well.

Given the attractiveness of pairing conformal prediction with prediction and classification methods, another natural extension would be conformal uncertainty sets for robust optimization with covariates. One could construct $\mathcal{U}$ as a function of a vector of covariates $x$,

\begin{equation}
    \mathcal{U}(x) = \{c \in \mathbb{R}^d: ||\hat{L}^{-}(u - \hat{u})||_2 \le \Omega_{\alpha}\},
\end{equation}

\noindent
where $\hat{u}$ is the prediction for the response vector $u$ given $x$ and $\hat{L}^-$ is an estimate of the lower triangular matrix of the Cholesky decomposition for the inverse-covariance matrix of $u$. While conformal uncertainty sets generated with covariate information would still be valid marginally, additional consideration could be taken to generate locally valid prediction regions and uncertainty sets. Locally valid prediction regions have recently been explored in \cite{diquigiovanni2021conformal} with respect to functional data through the use of a modulation function. 

\bibliography{master_bib}

\clearpage
\section{Supplementary Materials}

\begin{figure}[h]
    \centering
    \makebox[\textwidth][c]{
    \includegraphics[scale = 0.43]{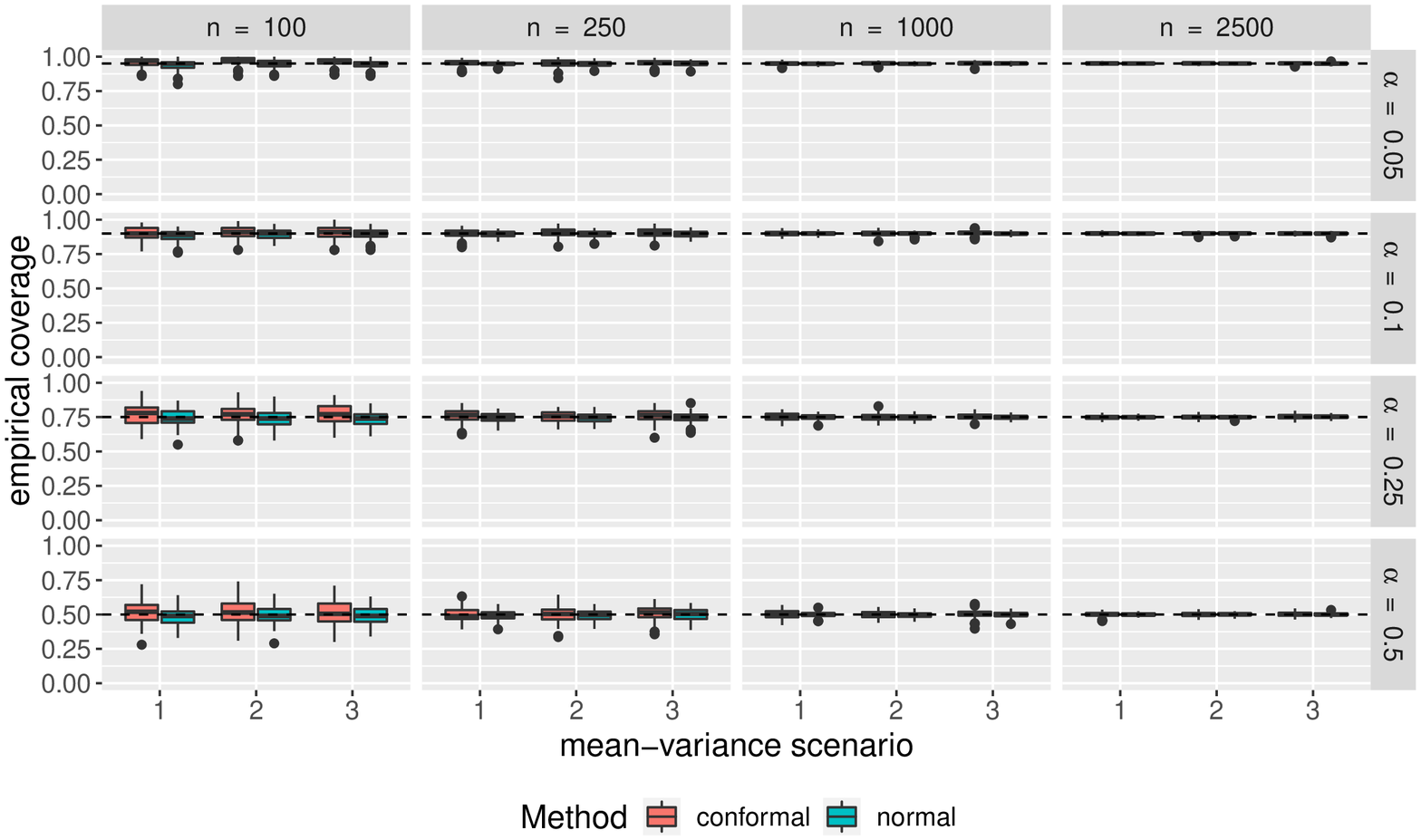}%
    \includegraphics[scale = 0.43]{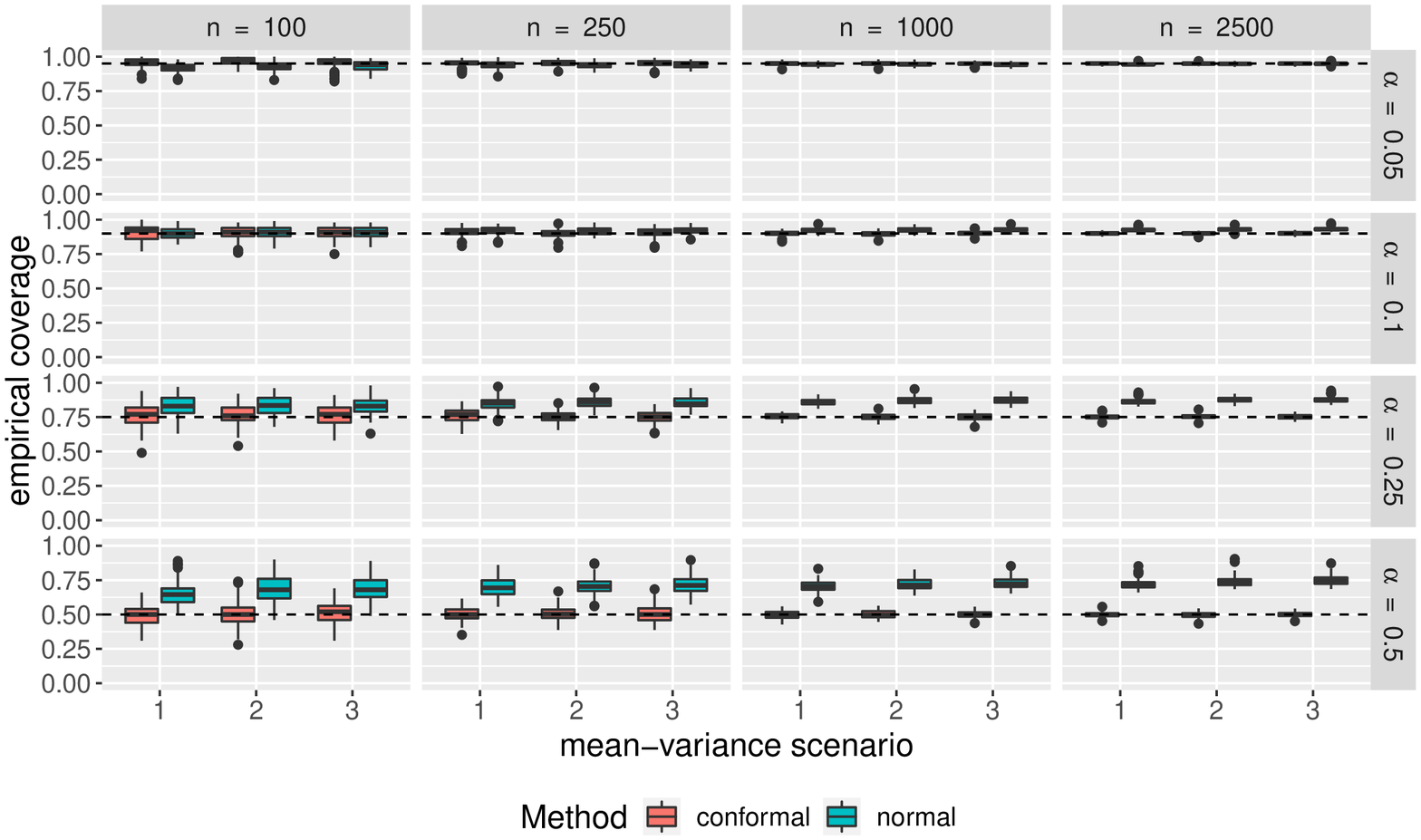}
    }
    \caption{Empirical coverage across $n$ test cases with 100 experiment repetitions for normally distributed cost (left) $t$-distributed cost (right) with $d = 2$.}
    \label{fig:covd2}
\end{figure}

\begin{figure}[h]
    \centering
    \makebox[\textwidth][c]{
    \includegraphics[scale = 0.43]{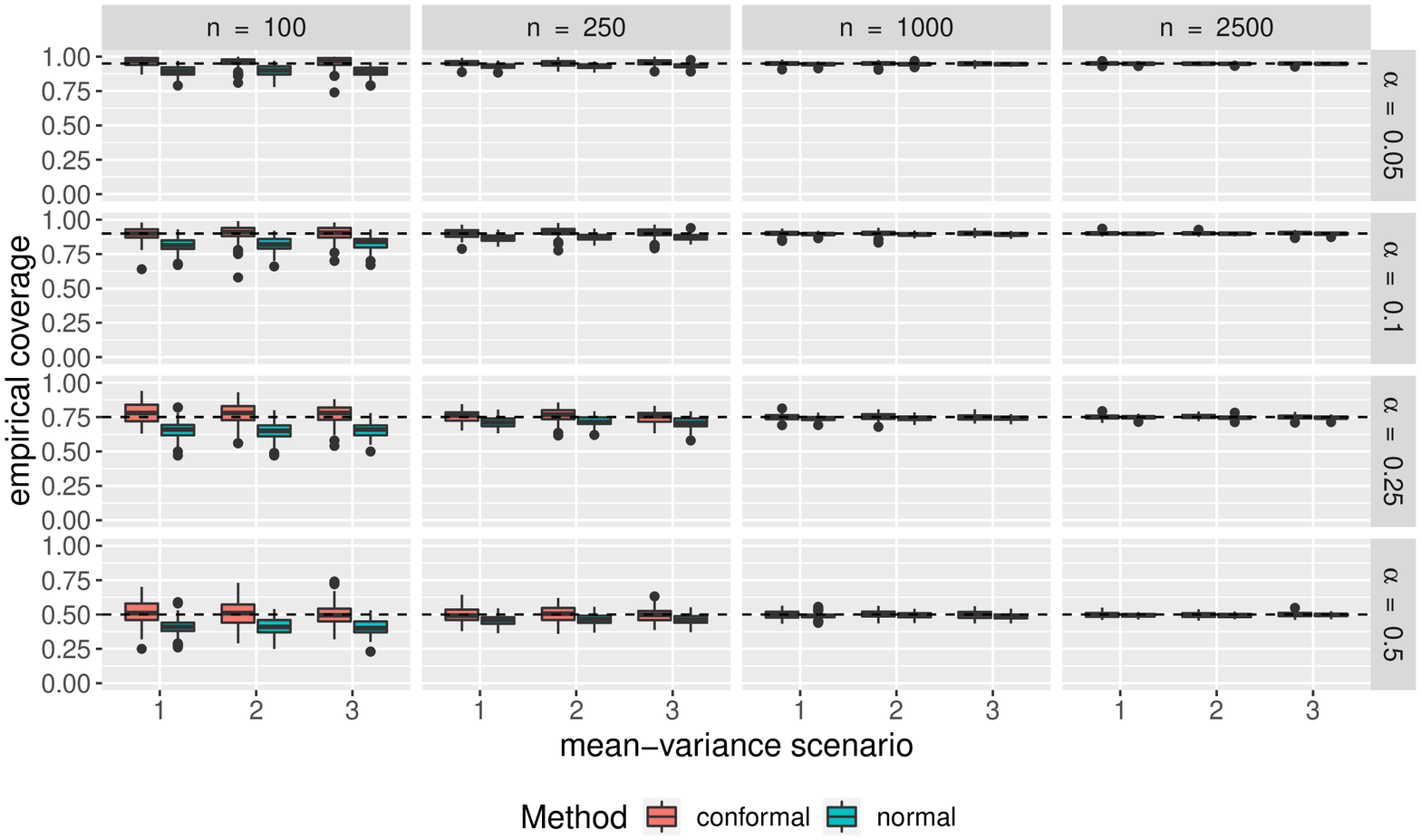}%
    \includegraphics[scale = 0.43]{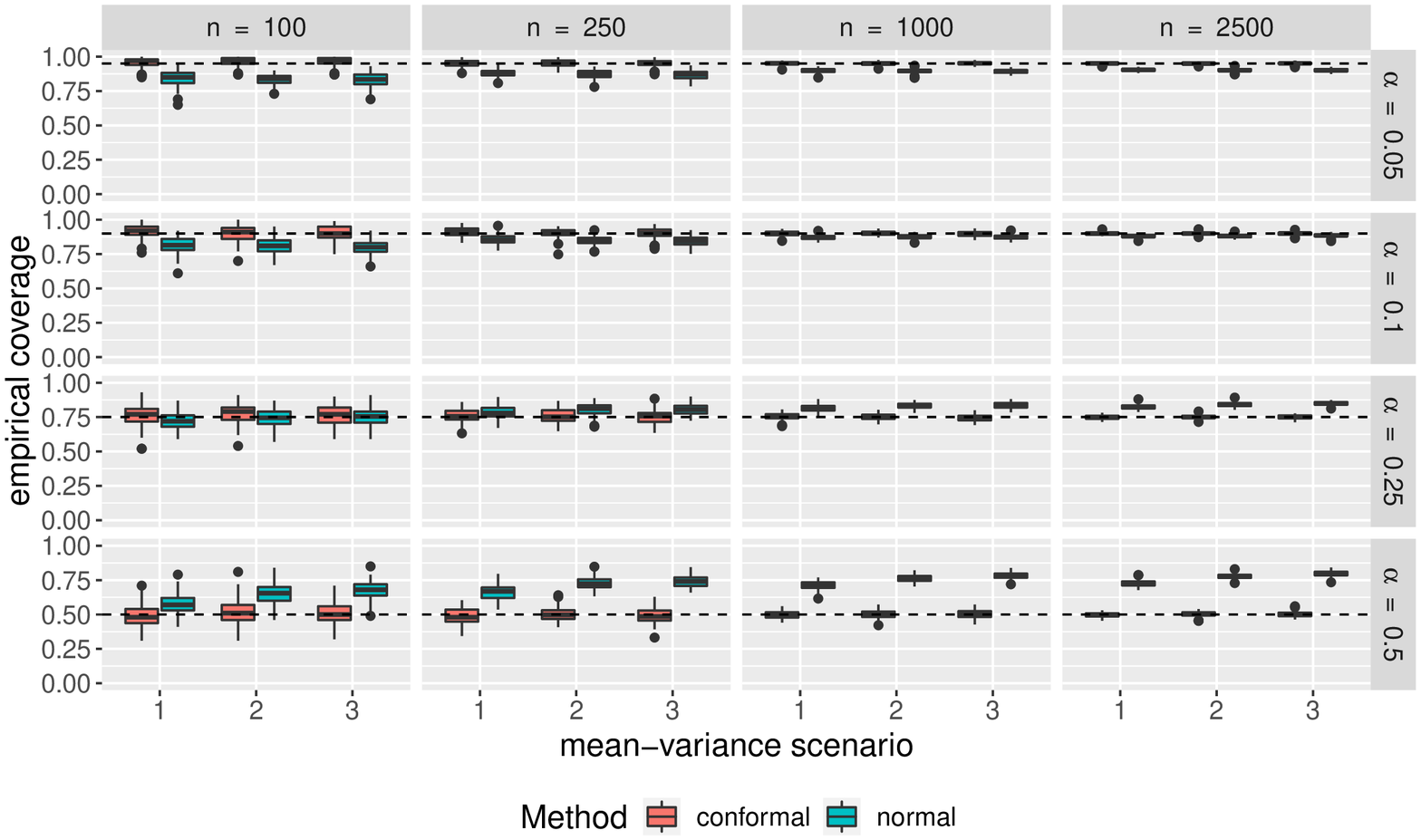}
    }
    \caption{Empirical coverage across $n$ test cases with 100 experiment repetitions for normally distributed cost (left) $t$-distributed cost (right) with $d = 10$.}
    \label{fig:covd10}
\end{figure}

\clearpage
\begin{figure}
    \centering
    \makebox[\textwidth][c]{
    \includegraphics[scale = 0.43]{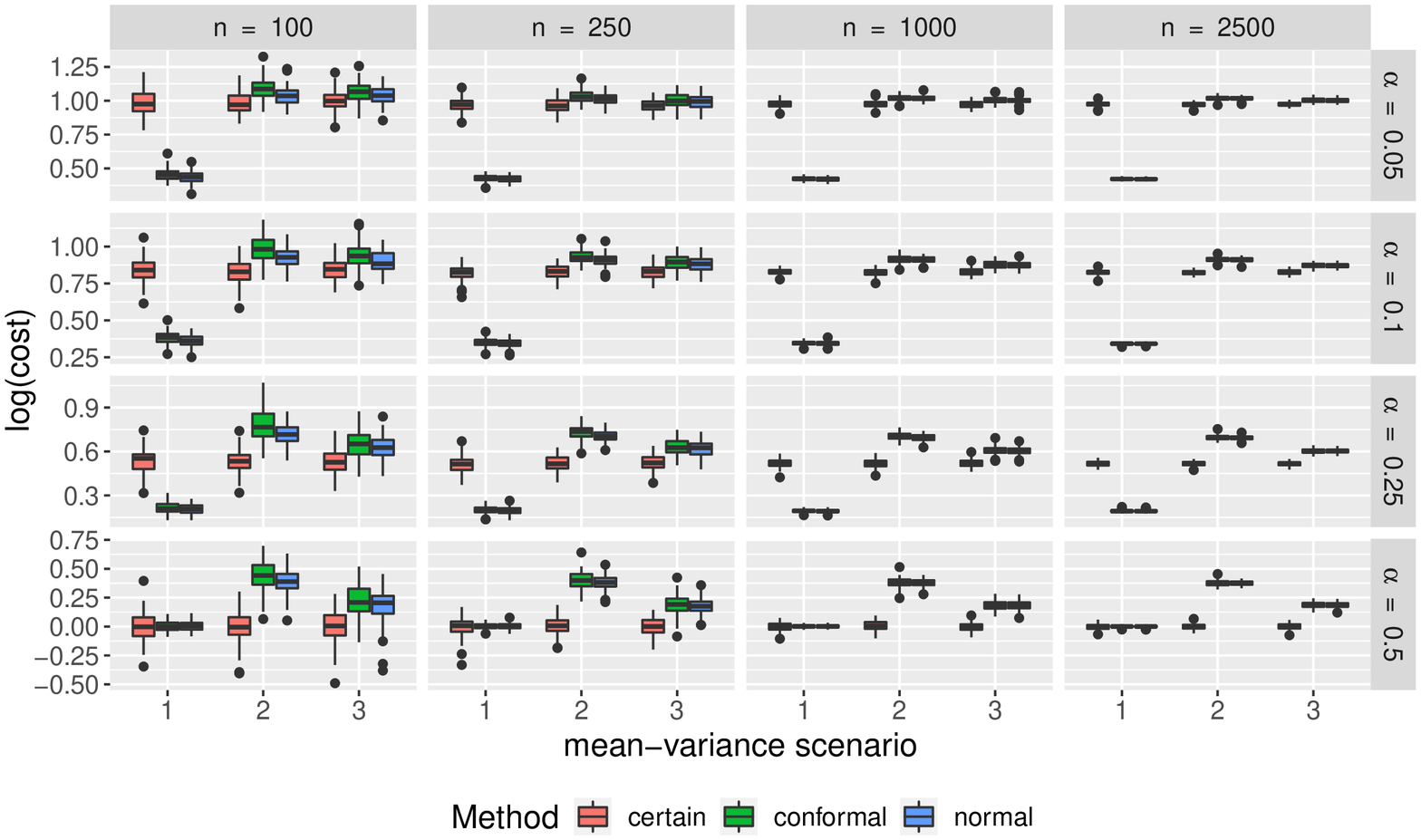}%
    \includegraphics[scale = 0.43]{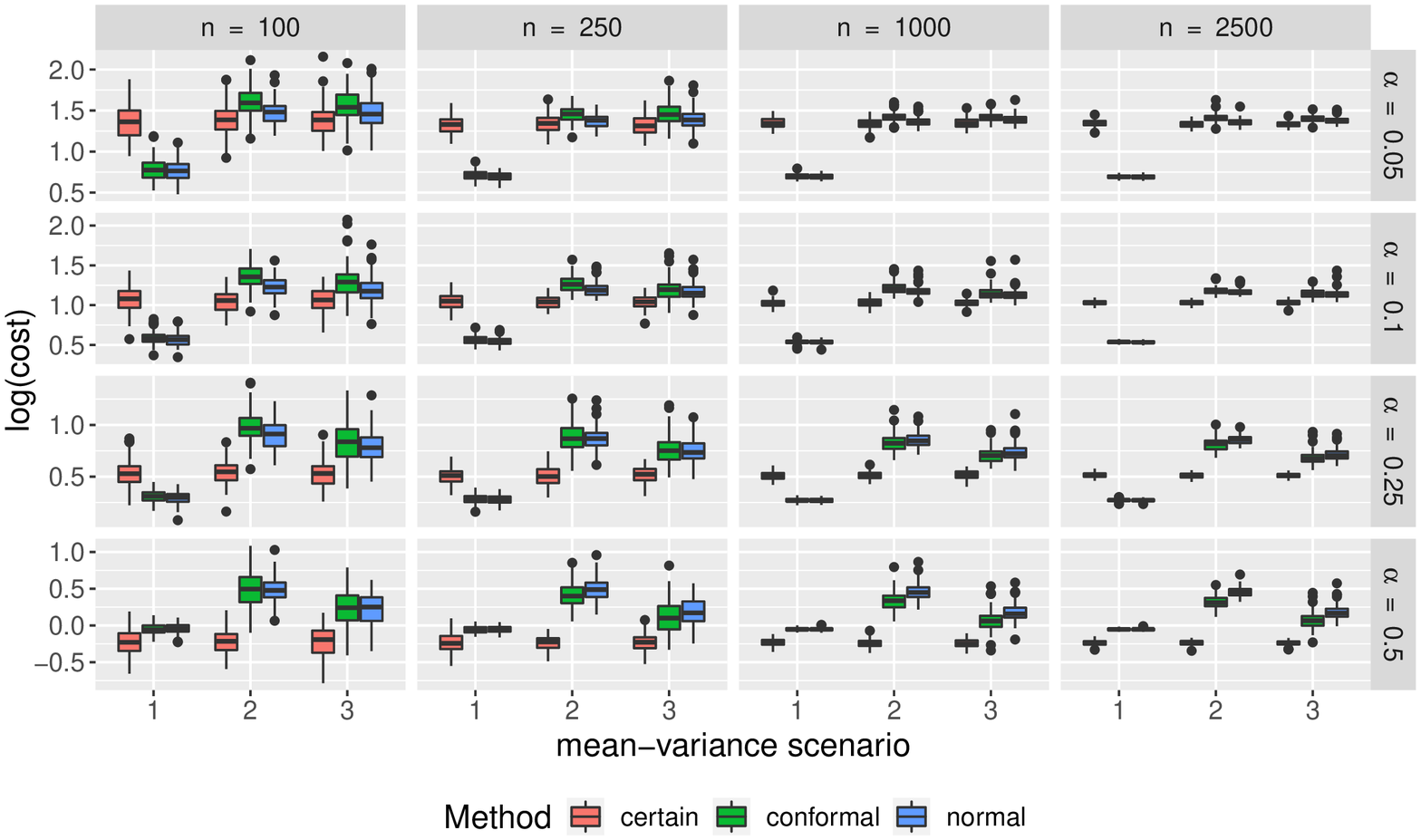}
    }
    \caption{$\log$ of $\alpha$ worst-case objective function value across $n$ test cases with 100 experiment repetitions for normally distributed cost (left) and $t$-distributed cost (right) with $d = 10$.}
    \label{fig:costd10}
\end{figure}

\end{document}